\documentclass[aps,pra,twocolumn,reprint,amsmath,amssymb,graphicx]{revtex4}

\newcommand{\beq}{\begin{equation}}
\newcommand{\eeq}{\end{equation}}
\newcommand{\bea}{\begin{eqnarray}}
\newcommand{\eea}{\end{eqnarray}}
\providecommand{\abs}[1]{\left\lvert#1\right\rvert}

\providecommand{\bra}[1]{\langle #1 \rvert}
\providecommand{\ket}[1]{\lvert #1 \rangle}
\providecommand{\bbra}[1]{\langle #1 \rvert\rvert}
\providecommand{\kket}[1]{\lvert\lvert #1 \rangle}

\newcommand{\ud}{\mathrm{d}}

\newcommand{\un}{\openone}

\usepackage{tikz}
\usepgflibrary{arrows}
\usetikzlibrary{decorations}
\usetikzlibrary{decorations.pathmorphing,patterns}
\usetikzlibrary{scopes}
\usetikzlibrary{shapes, matrix}
\begin{document}

\title{On the Ubiquity of Beutler-Fano Profiles: from Scattering to Dissipative Processes}

\author{Daniel Finkelstein-Shapiro \thanks{daniel.finkelstein_shapiro@chemphys.lu.se}}
\affiliation{Division of Chemical Physics and Nanolund, Lund University, Box 124, 221 00 Lund, Sweden}
\email[Corresponding author:$\;$]{daniel.finkelstein_shapiro@chemphys.lu.se}
\author{Arne Keller}
\affiliation{Laboratoire Mat\'eriaux et Ph\'enom\`enes Quantique,
B\^atiment Condorcet
10, rue Alice Domon et Leonie Duquet
75205 Paris cedex 13, CNRS-Univ. Paris-Diderot.  Univ. Paris-Sud, Univ. Paris-Saclay, France}

\begin{abstract}
Fano models - consisting of a Hamiltonian with discrete-continuous spectrum - are one of the basic toy models in spectroscopy. They have been successful in explaining the lineshape of experiments in atomic physics and condensed matter. These models however have largely been out of the scope of dissipative dynamics, with ony a handful of works considering the effect of a thermal bath. Yet in nanostructures and condensed matter systems, dissipation strongly modulates the dynamics. In this article, we present an overview of the theory of Fano interferences coupled to a thermal bath and compare them to the scattering formalism. We provide the solution to any discrete-continuous Hamiltonian structure within the wideband approximation coupled to a Markovian bath. In doing so, we update the toy models that have been available for unitary evolution since the 1960s. We find that the Fano lineshape is preserved as long as we allow a rescaling of the parameters, and an additional Lorentzian contribution that reflects the destruction of the interference by dephasings. The universality of the lineshape can be traced back to specific properties of the effective Liouvillian.
\end{abstract}

\maketitle

\section{Introduction}

The first observations of the distinctive asymmetric Fano profile came during the 1930s in the study of molecular photodissociation and atomic photoionization spectra \cite{Brown1932,Beutler1935}. Spurred by the new unexplained experimental evidence, O.K. Rice on the one hand \cite{Rice1933} and Ugo Fano on the other \cite{Fano1935} set out to develop the corresponding theories. The inclusion of pre-dissociated for the first and auto-ionizing states for the second opened an additional pathway towards the fragmentation. The interference pattern between the direct and the (newly included) indirect pathway resulted in the asymmetric lineshape. Although both theories had similar physics, eventually the simplicity of the Fano expression \cite{Fano1961} led to it being more widely used. The Fano profile:
\begin{equation}
f(\epsilon,q)=\frac{(\epsilon+q)^2}{\epsilon^2+1}
\label{eq:Fano}
\end{equation}
where $\epsilon$ is a normalized detuning of the laser frequency with respect to the ground-excited transition energy and $q$ is the lineshape asymmetry parameter that reflects the relative strength of the two pathways to reach the continuum.
The theory was generalized to include more complicated and realistic structures of the energy levels whilst preserving the general idea of competing pathways to a continuum \cite{Fano1961,Glutsch2002}.
Over the course of the years, the Fano profile has been successful in fitting the lineshape in situations far removed from photodissociation or photoionization of molecular or atomic gases experiments \cite{Miroshnichenko2010,Lucky2010,
Nguyen2007,Lombardi2016,Gallinet2010, Harbers2007,Brown2001,Mazumdar2006,Tremblay2010,Bianconi2003,Vittorini-Orgeas2009}. \newline



Descriptions of asymmetric lineshapes in condensed matter systems followed very shortly after 1961 \cite{Baldini1962,Jain1965}. These were found in GaAs semiconductors under a magnetic field, where magneto-excitons resulting from excitation of Landau levels couple Coulombically to the band continuum \cite{Glutsch1994,Siegner1995,Siegner1995a,Siegner1996,Seisyan2016}. Holfeldt et al. measured Fano resonances in biased superlattices where Wannier-Stark excitons coupled to the continuum of higher transitions \cite{Holfeld1998}.
More recently Fano lineshapes in optical excitations coupled to narrow phonon modes have been reported \cite{Yoshino2015}. The reports of Fano asymmetries in metamaterials have been even more prolific \cite{Miroshnichenko2010,Lucky2010}. There is a particular technological interest in these metamaterials. They serve as very high enhancement substrates for SERS and also as efficient nonlinear media. \newline

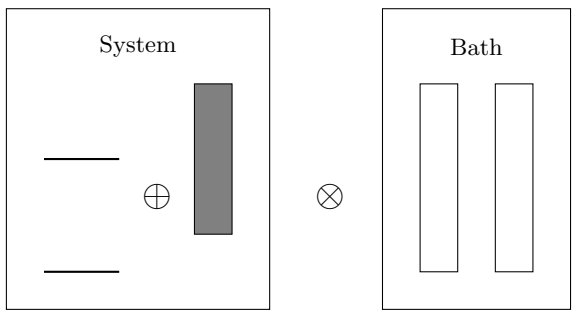
\begin{figure}[ht]
\centering
\begin{tikzpicture}[scale=0.5]
\draw[thick] (0,0) -- (2cm,0);
\draw[thick] (0,3cm)--(2cm,3cm);
\draw[fill=gray]  (4cm,1cm) rectangle (5cm,5cm);
\draw  (-1cm,-1cm) rectangle (6cm,7cm);
\draw  (9cm,-1cm) rectangle (14cm,7cm);
\draw  (10cm,0cm) rectangle (11cm,5cm);
\draw  (12cm,0cm) rectangle (13cm,5cm);
\node at (2.5,6) {System};
\node at (11.5,6) {Bath};
\node at (3,2) {\Large $\oplus$};
\node at (7.6,2) {\Large $\otimes$};
\end{tikzpicture}
\caption{\label{fig:FanoDissipRes} Energy structure of the Fano model with dissipation. It is important to stress that there are two continua: the one belonging to the system which belongs to the same particle as the dicrete states which are all coupled to a continuum of bosonic modes conforming the bath. The mathematical operations of direct sum and tensor product emphasize this distinction.}
\end{figure}

The theoretical problem of including the coupling to a thermal bath was recognized without an explicit solution by Ugo Fano in 1963, inspired by the problem of pressure broadening \cite{Fano1963}. The motivation is as follows: the system consisting of a discrete and continuum manifolds is coupled to another continuum of modes of the bath that can exchange energy with the system (see Figure \ref{fig:FanoDissipRes}). It is important to stress that there are two continua which are qualitatively different. One corresponds to the system one-particle states, and the other one to excitations of others particles (for example modes of a phonon bath). 
Rzazewski and Eberly considered the effect of phase incoherence of the incoming laser at arbitrary strengths of the field and solved exactly the energy resolved population of the continuum using stochastic methods \cite{Rzazewski1983}. In a landmark work, Agarwal and co-workers considered the case of a Fano model coupled to the vacuum modes of light to account for spontaneous emission \cite{Agarwal1982,Agarwal1984}. They provide the exact solution in the case of weak field, and compact expressions for the arbitrary-field strength case (within the rotating wave approximation) in terms of the poles of the resolvent for the continuum populations. The main conclusions unveiled by the expressions is that spontaneous-emission induced decay preserves the Fano minima in both the weak and strong field cases, and the authors examine the consequences on the photoelectron spectra, in particular the effect of electron recycling and the presence of sink states. This is explored further in subsequent work \cite{Agassi1984, Haan1987, Ravi1991}. The role of radiation damping was investigated through the one and two-time correlation function of the electric dipoles by Haus et al. \cite{Haus1983}.

More recent efforts have considered the dissipation in the context of condensed matter instead of ionization. Zhang et al. in a series of papers solved for the absorption of a system with Lindblad dissipator in the strong dissipation regime \cite{Zhang2006,Zhang2011}. They predicted a diminishing asymmetry with increasing field-intensity, which has been verified experimentally and called the nonlinear Fano effect \cite{Kroner2008}. Gallinet and co-workers introduced an analytical continuation of the Fano form and captured the effects of dissipation in plasmonic devices \cite{Gallinet2010}. The solution of scattering by a lossy dielectric has been solved by Tribelsky and Miroshnichenko more recently \cite{Tribelsky2016}. Barnthaller et al. followed a similar treatment in their theory/experiment study of dissipation in waveguides \cite{Baernthaler2010}. Finkelstein-Shapiro et al. calculated the weak-field emission of a Fano model with a Lindblad dissipator and found that the parameters of the lineshape are rescaled by the dissipation \cite{Finkelstein2015}. They then provided the exact solution of the Fano model in the wideband approximation coupled to a Markovian bath \cite{Finkelstein2016-1}.

It is important to recognize that the Fano lineshape can refer to many types of interference: as in the original problem, the interference of quantum mechanical amplitudes in a scattering experiment \cite{Rau2004}, but also the interference of light modes, but even of the interference present in driven coupled classical oscillators \cite{Joe2006,Satpathy2012}. In each of these cases the dissipation is added differently. The interference of quantum mechanical amplitudes is the most challenging case due to the details of the system-bath coupling, restrictions on the complete positivity of the density matrix and situations of non-Markovianity \cite{Breuer2004}. We focus our attention in this article on this case. 
It is relevant because 
it is still not known how to systematicaly deal with a discrete-continuum Hamiltonian coupled to a Markovian reservoir.
The scattering solution arises from solving the Schrodinger equation in Hilbert space while the dissipative solution arises from solving the Liouville equation in Liouville space, yet experiments pertaining to both cases are equally fit by this expression, lending a strong character of universality to this lineshape.
 \newline

In this article, we aim to provide a clear context for the theoretical problem of Fano interferences in dissipative Liouville space and formalize the origins of its wide applicability mathematically. In part II we review the solution of the scattering Hilbert space problem using Feshbach projection and resolvents. In part III the solution for the dissipative Lindblad dynamics is given using the same methods but in the space of superoperators. The mathematical condition to obtain a Fano profile is made evident and we provide a systematic procedure and recipe to deal with any discrete-continuum Hamiltonian under the wideband approximation. We discuss how to apply the equations to a transport process.

\section{Scattering formalism}
\begin{figure}
\includegraphics[width=0.4\textwidth]{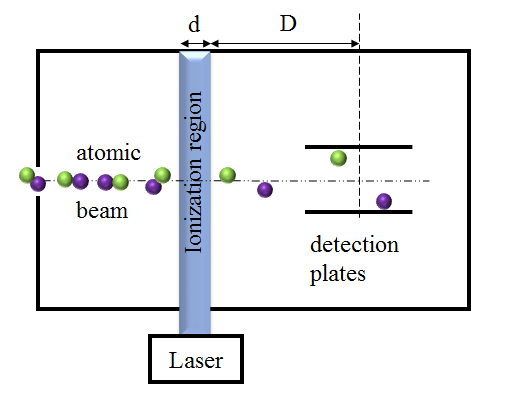}
\caption{Scattering experiment: an atomic or molecular beam traverses an interaction region with a laser of variable length and ionizes. The fragmented ionized species are detected at Faradaic plates.}
\label{fig:scattering_experiment}
\end{figure}

We consider the typical photoionization experiment (Figure \ref{fig:scattering_experiment}).
The atoms or molecules pass through an interaction region with a laser beam where they become ionized. The charged fragments are then detected capacitevely. The light-matter interaction region $d$ corresponds to an interaction time $T$ which depends upon the speed of the molecular or atomic beam. Detector plates are placed at a distance $D$ from the beam. The Hamiltonian corresponds to:
\begin{align}
&H=H_0+H_V+H_F \\
&H_0=E_{0}\ket{g}\bra{g}+E_e\ket{e}\bra{e}+\int dk E_k\ket{k}\bra{k} \nonumber \\
&H_V=\int dk \big[ V(k)\ket{e}\bra{k}+V(k)^*\ket{k}\bra{e} \big] \nonumber \\
&H_F=F \left[\mu_{e}\cos(\omega_L t)\ket{g}\bra{e}+\mu_{e}^*\cos(\omega_L t)\ket{e}\bra{g}\right] \nonumber \\
&+F \int dk \left[\mu_{c}(k)\cos(\omega_L t)\ket{g}\bra{k}+\mu_{c}^*(k)\cos(\omega_L t)\ket{k}\bra{g}\right],
\label{eq:Hamiltonian}
\end{align}
where $H_0$ is a bare Hamiltonian, $H_V$ is the coupling of the discrete excited state $\ket{e}$ to the continuum set of states $\ket{k}$ and $H_F$ is the interaction with the incident radiation field of amplitude $F$ and angular frequency $\omega_L$, allowing transitions from the ground state $\ket{g}$ to the discrete excited state with transition dipole moment $\mu_e$ and to the continuum of states with transition dipole moment $\mu_c$. Without loss of generality we take $V,\mu_e,\mu_c$ to be real. We solve the dynamics in a rotating frame obtained by applying the unitary transformation $U_L(t) = e^{-i\Omega_L t}$, where $\Omega_L = \omega_L\ket{g}\bra{g}$,  so that in the new frame the Hamiltonian becomes $\hbar\Omega_L + U_LHU_L^{-1}$. Within  the rotating wave approximation (RWA), which consists  in neglecting fast oscillating terms, this Hamiltonian in the rotating frame is time independent. To obtain explicit results, we will make use of the wideband approximation  which consists in neglecting the $k$-dependence for the coupling $V(k)$ and $\mu_c(k)$ in the Hamiltonian~Eq.~\eqref{eq:Hamiltonian} and assuming a linear dispersion relation for the continuum $n=dk/dE$.

The solution of the system's dynamics is fully specified by the evolution operator $U(t)$:
\begin{equation}
\Psi(t)=U(t)\Psi(0).
\end{equation}
If the detection plates are far enough from the ionization region so that all the excited atoms have had time to completely ionize, then the ionization probability  is given by \cite{Lambropoulos1981}:
\begin{equation}
P(T)=1-\abs{U_{gg}(T)}^2,
\end{equation}
where $\abs{U_{gg}(T)}^2$ is the probability of finding the system in the ground state at time $T$ assuming it has started in the ground state.
We use the notation  $A_{ij}=\bra{i}A\ket{j}$ for an operator $A$. We have:
\begin{equation}
\label{eq:fourierLaplace}
U_{gg}(t)=-\frac{1}{2\pi i} \int_{\mathbb{R}+i\eta} G_{gg}(z)e^{-izt/\hbar}\ud z,
\end{equation}
where $\eta$ is any positive number and $G(z)=(z-H)^{-1}$ is the Hamiltonian resolvent, that can also be written as:
\[
G(z) = -\frac{i}{\hbar} \int_{0}^{+\infty}U(t)e^{izt/\hbar} \ud t \text{ for } \Im[z]>0.
\]
 Lambropoulos and Zoller solved for the scattering cross-section under intense fields using the resolvent approach \cite{Lambropoulos1981}. Here we show how the result is obtained using the resolvent approach along with projection operators \cite{Feshbach1962}. We will build on this result in later sections to solve the dissipative case of a Fano system coupled to a thermal bath.
Projection operators allow us to separate the subspace that corresponds to the discrete states from the subspace that corresonds the continuum. These are:
\begin{equation}
P=\ket{g}\bra{g}+\ket{e}\bra{e}; \quad
Q=\int_{-\infty}^{\infty} dk \ket{k}\bra{k}.
\label{eq:PQdef}
\end{equation}
Separating  the Hilbert space in this way allows us to write the the restriction $PG(z)P$ of the exact resolvent $G(z)$ in the discrete subspace as $PG(z)P=(z-H_{\text{eff}})^{-1}$, where $H_{\text{eff}}$ is an effective Hamiltonian acting on the discrete subspace only, but incorporating the effect of the continuum. It is given by $H_{\text{eff}}=PHP+PHQG_0(z)QHP$, where $G_0$ is the resolvent of $H_0$, $G_0(z) = (z-H_0)^{-1}$.
 Because the problem only involves the evolution operator element that evolves the wavefunction from the ground state onto the ground state at a later time, the solution is fully specified in the $P$ subspace (See Appendix~\ref{app:Proj}). \newline

We switch to dimensionless  variables where energies are given in units of $\hbar\gamma = n\pi V^2$, where $n$ is the density of states $n = \frac{\ud k}{\ud E}$, and times are given in units of $1/\gamma$. In the wide band approximation where $V(k)$ end $\mu_c(k)$ are taken as constants independent of $k$,
the integration over $k$ is explicit and
the effective Hamiltonian in the RWA approximation
introduced in the previous section is:
\begin{equation}
H_{\text{eff}}=\begin{bmatrix}
-i\Omega^2 & \Omega(q-i) \\
\Omega(q-i) & -\epsilon-i
\end{bmatrix},
\label{eq:Heff}
\end{equation}
where $q=\frac{\mu_e}{n\pi \mu_c}$, $\epsilon=(\omega_L-E_e)/\hbar\gamma$, $\Omega=\mu_cF/2V$. The matrix element $G_{gg}(z)$ of the resolvent is given by:
\begin{equation}
G_{gg}(z)=\frac{z+\epsilon+i}{(z-z_1)(z-z_2)},
\end{equation}
where
\begin{equation}
\begin{split}
z_{1,2}&=-\frac{1}{2}(\omega_0\pm\omega), \text{ with }
\omega_0=\left(\epsilon+i(1+\Omega^2)\right) \\
\text{with }  \omega&=\left\{(\epsilon+i)^2 -2\Omega^2\left[1-2q^2 +i(4q + \epsilon)\right] -\Omega^4\right\}^{\frac{1}{2}}.
\end{split}
\end{equation}
We stress that $\omega_0$ and $\omega$ are complex numbers and for $\omega$ both determination of the square root can be used as this choice only affect the conventional labeling of $z_1$ and $z_2$. The inverse Fourier Laplace Eq.~\eqref{eq:fourierLaplace} is immediately obtained and gives:
\begin{align}
U_{gg}(t) &= a_1e^{-iz_1 t} + a_2e^{-iz_2 t}, \\
\text{with } a_1 &= \frac{z_1+\epsilon+i}{z_1-z_2}; \quad a_2 = \frac{z_2+\epsilon+i}{z_2-z_1}. \nonumber
\label{eq:Ugg}
\end{align}
The probability $\abs{U_{gg}(t)}^2$ to remain in the ground state at time $t$, can be written as~:
\beq
\begin{split}
\abs{U_{gg}(t)}^2 &= e^{-(1+\Omega^2) t}\left(\abs{a_1}^2 e^{\Im[\omega] t } +
\abs{a_2}^2 e^{-\Im[\omega ]t } \right.  \\
&+ \left. 2\Re\left[a_1a_2^* \right] e^{-i\Re[\omega] t} \right) \\
&=\abs{a_1}^2e^{-\Gamma_0 t} + \abs{a_2}^2e^{-\Gamma_2 t}\\
&+2\Re[a_1a_2^*]e^{-i\Re[\omega]t}e^{-\Gamma_1 t}
\end{split}
\eeq
\begin{figure}
\includegraphics[width=0.5\textwidth]{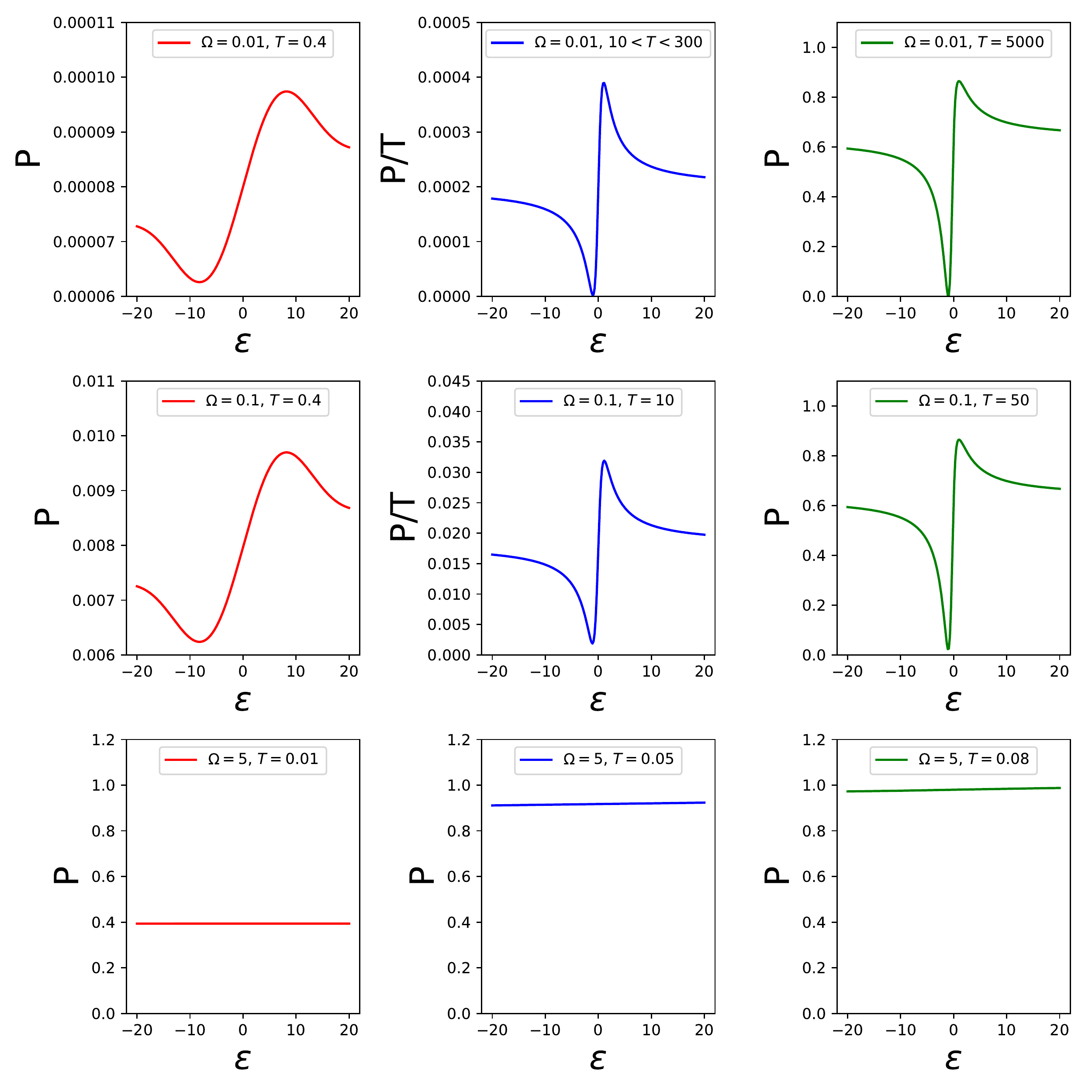}
\caption{Ionization probability for $q=1$ for three values of the field $\Omega=0.01$ (first row), $\Omega=0.1$ (second row) and $\Omega=5$ (last row). For each field intensity three times are shown. For weak fields ($\Omega=0.01$ first row), during the early times (on the order of $1/n\pi V^2$) the interference pattern is building up. The steady-state rate (middle column, in blue) spans a large time window (the rate $P/T$ does not show a significant change for a time window between $T=10$ and $T=300$). At large times the profile tends towards a Fano-like shape with smaller $q$ parameter although the Fano equation is no longer exact. At intermediate field $\Omega=0.1$, second row, we observe a similar behavior although the time window in which the Fano equation is valid is much smaller. At strong fields $\Omega=5$, last row,  the profile is flat and increases with time until the species have fully fragmented. All time values are given in units of $1/n\pi V^2$.}
\label{fig:scattering}
\end{figure}
%
To recover the original Fano fragmentation rate, we first have to  consider a  situation where the concept of a rate has a meaning. This is the case if there is a decoupling between the different timescales:
$\Gamma_0 = 1+\Omega^2 -\Im{(\omega)}$, $\Gamma_1 = 1+\Omega^2$  and  $\Gamma_2 = 1+\Omega^2 +\Im{(\omega)}$. This decoupling occurs in the low field limit $\Omega^2 \ll 1$. Indeed
in this limit we have~:
\begin{align*}
\Gamma_0  &= 2\Omega^2\frac{\left(\epsilon + q\right)^2}{1+\epsilon^2} + \mathcal{O}(\Omega^4); \\
\Gamma_2 &=  2+2\Omega^2\left[1 - \frac{\left(\epsilon + q\right)^2}{1+\epsilon^2}\right]+\mathcal{O}(\Omega^4)
\end{align*}
and therefore, in this limit $\Gamma_0 \ll \Gamma_1\simeq \Gamma_2$. Hence, in the range $1/\Gamma_2 \ll t \ll1/\Gamma_0$, $P(t)$ can be written as~:
\[
P(t) = 1-\abs{U_{gg}}^2 \simeq 1-\abs{a_1}^2e^{-\Gamma_0 t}\simeq 1-\abs{a_1}^2\left[1-\Gamma_0t \right].
\]
Furthermore, at the same order of approximation, $\abs{a_1}^2 = 1 - \frac{\Omega^2}{\epsilon^2+1} +\mathcal{O}(\Omega^4)$. Consequently,
\[
P(t) = \frac{\Omega^2}{\epsilon^2+1} +2\Omega^2\frac{(\epsilon +q)^2}{1+\epsilon^2}t +\mathcal{O}(\Omega^4).
\]
The fragmentation rate is therefore given by:
\begin{equation}
\frac{\ud P}{\ud t} = 2\Omega^2\frac{(\epsilon +q)^2}{1+\epsilon^2} + \mathcal{O}(\Omega^4),
\label{eq:scattRate}
\end{equation}
which is indeed proportional to the Fano profile.
In the experiment described in Figure \ref{fig:scattering_experiment}, a molecular or  atomic beam with constant flux goes through an ionization region of duration $T$. The total number of detected ions is:
\begin{equation}
\begin{split}
N_{\text{ionized}}&=\text{flux}\times P(T) \\
&\approx \text{flux} \times T\frac{dP(t)}{dt}|_{t=0^+} = 2 T\Omega^2\frac{(\epsilon +q)^2}{1+\epsilon^2},
\end{split}
\end{equation}
where the second line is an approximation that gives an exact Fano profile and $t=0+$ means that we consider a sufficiently  small time  such that $t \ll \frac{1}{\Gamma_0}$ which is realized if $t\ll\frac{1}{(1+q^2)\Omega^2}$,
but not too small in the sense that $t\gg 1$ (in units of $\hbar/n\pi V^2$). This is possible for all values of $\epsilon$, only if $\Omega^2(1+q^2)\ll 1$.
It is in this specific sense that  we can say that
for short interaction times but much bigger than $1$, at weak fields, we measure the Fano profile. The need for a short elapsed time for the interference to build-up is not usually recognized, but has been measured experimentally \cite{Kaldun2016}.

Figure \ref{fig:scattering} shows the evolution of the ionization profile at different times (in units of $1/n\pi V^2)$ for an asymmetry parameter of $q=1$ for three values of the field $\Omega=0.01$, $\Omega=0.1$ and $\Omega=5$. For weak fields ($\Omega=0.01$ and $\Omega=0.1$), the Fano profile does not appear right away but builds up during a time $1/n\pi V^2$. This transient corresponding to the sampling time of the continuum by the discrete excited state has been observed experimentally \cite{Kaldun2016}. After this time the Fano expression then develops and is valid for $T \ll 1/\Gamma_0$. At long times, there is a saturation effect where the profile is Fano like with a smaller asymmetry parameter, although we note that this lineshape cannot even approximately be described by a Fano form. This evolution of the profile is absent for strong fields ($\Omega=5$) which shows a flat profile at all times.

To conclude this section we would like to stress that the explanation for the asymmetric profile originally given by Fano is valid for  very specific experimental conditions, namely scattering with weak field and observations at intermediate times, which were well adapted to the experimental setup available at that time. In this type of scattering experiment, the  incident flux of particles is constant. However, the Beutler-Fano formula has been used in a plethora of experimental contexts where these specific conditions are not fulfilled.  One of our objective in this article is to understand if and why the Beutler-Fano formula works in these variety of context.
Surprisingly, as we show in the next section, taking into account dissipative processes broadens the conditions under which Beutler-Fano profiles can be observed, justifies its choice as a phenomenological fit and explains its overwhelming success.

\section{Dissipative formalism}
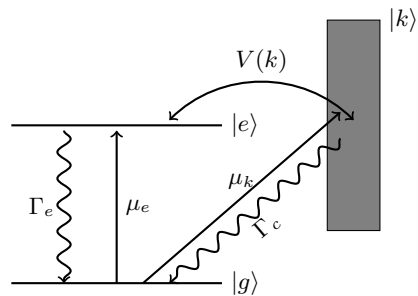
\begin{figure}[ht]
\centering
\begin{tikzpicture}[scale=0.7]
\draw[thick] (0,0) -- (4cm,0) node[right]{$\ket{g}$};
\draw[thick] (0,3cm)--(4cm,3cm) node[right]{$\ket{e}$};
\draw[fill=gray]  (6cm,1cm) rectangle (7cm,5cm) node[right]{$\ket{k}$} ;
\draw[->,thick] (2cm,0) --(2cm,2.9cm) node[midway, right] {$\mu_{e}$};
\draw [->,thick] (2.5cm,0cm)--(6.25cm,3.25cm)
node[midway, above] {$\mu_{k}$};
\draw[<->,thick] (3cm,3.1cm) to[out=45,in=135] node [sloped, above] {$V(k)$} (6.5cm,3.1cm);
\draw [->,thick,decorate,decoration=snake] (6.25cm,2.75cm)--(3cm,0)
node[midway,sloped ,below,] {$\Gamma_c$};
\draw [->,thick,decorate,decoration=snake] (1cm,2.9cm)--(1cm,0) node[midway,left] {$\Gamma_e$};
\end{tikzpicture}
\caption{\label{fig:FanoDissip} Energy levels and transitions of a Fano-type model with dissipation. Hamiltonian coupling are indicated by straight arrows, dissipative processes by twisted arrows. Population relaxations from continuum states, at a rate $\Gamma_c$, and from the discrete excited state, at a rate $\Gamma_e$ to the ground state are only considered.}
\end{figure}

%
Let consider a steady-state monochromatic irradiation that  impinges upon the system, as in the original Fano scattering case, but now we suppose that the   "fragmented" species (such as an electron in a continuum or in a conduction band) are restored to the ground state by dissipative processes induced by an environment. For instance, in gas phase  this environment can be collisions with others particles or  vacuum field fluctuation inducing spontaneous emission. In condensed phase, the environment may be constituted by phonons or by the Coulomb interaction of the photogenerated electron-hole pair.
In any case, the environment is here considered  as a Markovian bath at zero temperature. In such case, the evolution of the system state, represented by the density operator $\rho(t)$, fullfils a Liouville equation:
\begin{equation}
\frac{\ud \rho}{\ud t} = L(\rho)=-\frac{i}{\hbar}[H,\rho]+L^D(\rho),
\end{equation}
 where the generator of the dissipative evolution has the  Lindblad form,  $L^D=\sum_i \Gamma_i\big(  D_i\rho D_i^{\dagger}-\frac{1}{2}\{D_i^{\dagger}D_i,\rho \}\big)$ where $D_i$ are Krauss operators \cite{Lindblad1976,Gorini1976}.
We will repeatedly use the isomorphism from the column form to the tensorial product~\cite{Havel2003}
given by $L\tilde{\rho}R\rightarrow \bar{R} \otimes L \rho$, where $\rho$ is the
column form of $\tilde{\rho}$ through the correspondance:
$\tilde{\rho} = \sum_{ij} \rho_{ij}\ket{i}\bra{j} \rightarrow \rho =
\sum_{ij} \rho_{ij} \kket{ij}$, where $\kket{ij} \equiv \ket{i}\otimes\ket{j}$. We start by considering that the dissipation induces population relaxation from  the continuum states  to the ground state, and from the discrete excited state to the ground state, only (see Fig.~\ref{fig:FanoDissip}).

The explicit expression of $L$ has a Hamiltonian part $H$ that is given by Eq.~\eqref{eq:Hamiltonian},
and a dissipative part $L^D = L^D_{\text{pop}}+L^D_{\text{pure}}$.
Where $L^D_{\text{pop}}$ describe the relaxation of excited state populations.
Pure dephasing, in other words additional decay of coherence, is described by
$L^D_{\text{pure}}$.
\begin{align}
L^D_{\text{pop}}&=\int dk \Gamma(k) \Big\{ A(k,g)\otimes A(k,g)  \nonumber \\
& - \frac{1}{2}\left[1\otimes A^{\dagger}(k, g)A(k,g) +
  A^{\dagger}(k,g)A(k,g)\otimes 1\right]\Big\}   \nonumber \\
&+\Gamma_e \Big\{ A(e,g)\otimes A(e,g)  \nonumber \\
& - \frac{1}{2}\left[1\otimes A^{\dagger}(e, g)A(e,g) +
  A^{\dagger}(e,g)A(e,g)\otimes 1\right]\Big\},
\label{eq:dissipation_k}
\end{align}
\begin{eqnarray}
&L^D_{\text{pure}} =-\gamma_{eg}\big[\ket{e}\bra{e}\otimes\ket{g}\bra{g}+\ket{g}\bra{g}\otimes\ket{e}\bra{e}\big] \nonumber\\
&-\int dk\gamma_{kg}\big[\ket{k}\bra{k}\otimes\ket{g}\bra{g}+\ket{g}\bra{g}\otimes\ket{k}\bra{k}\big] \nonumber \\
&-\int
dk\gamma_{ke}\big[\ket{k}\bra{k}\otimes\ket{e}\bra{e}+\ket{e}\bra{e}\otimes\ket{k}\bra{k}\big],
\label{eq:pureD}
\end{eqnarray}
$A(i,j)=\ket{j}\bra{i}$ are the
jump operators and $\Gamma(k)$ is the population relaxation rate from
state $\ket{k}$ to $\ket{g}$ as is $\Gamma_e$ for the $\ket{e}$ population. $\gamma_{ij}$ is the pure dephasing rate for the $ij$ coherence.
As in the scattering case, we apply the RWA approximation, which gives a time-independent Liouvilian operator $L$ in the rotating frame, obtained through the unitary transformation
$L=e^{i\underline{\Omega}_Lt}\mathcal{L}(t)e^{-i\underline{\Omega}_Lt}$. Where $\mathcal{L}(t)$ is the original time-dependent  Liouville operator and $\underline{\Omega}_L$ is a diagonal matrix whose
elements are equal to  $\pm \omega_L$ for excited(ground)-ground(excited) coherences, and zero elsewhere.
That is
$\underline{\Omega}_L=\omega_L\left(\kket{eg}\bbra{eg} +\int \ud k \kket{kg}\bbra{kg}\right)
- \omega_L\left(\kket{ge}\bbra{ge}  +\int \ud k \kket{gk}\bbra{gk}\right)$.
The observable originally addressed by Fano is
 the total population in the continuum set of states under steady-state conditions and  we will focus in this observable.
The absorption cross-section $\Im(\rho_{eg}+\int dk \rho_{kg})$ can just as easily be calculated as both it and the total population are obtained from a knowledge of the steady-state density matrix.
\subsection{Feshbach projectors, resolvent and effective Liouvillian}
Our objective is to compute the steady state $\frac{\ud \rho}{\ud t} = 0$, which is the kernel of $L$, that is the solution of $L(\rho) = 0$.
In analogy to the scattering case, we define partition superoperators \cite{Finkelstein2016-1}:
\begin{equation}
\mathcal{P}=P\otimes P;\quad%
\mathcal{Q}=P\otimes Q + Q\otimes P + Q\otimes Q.
\label{eq:partitions}
\end{equation}
where $P$ and $Q$ have been defined in Eq.~\eqref{eq:PQdef}.
It can be shown that the complete kernel of $L$ can be obtained in two steps (see Appendix~\ref{app:Proj}).
\begin{equation}
\begin{split}
L_{\text{eff}}\mathcal{P}\rho&=0 \\
\mathcal{Q}\rho&=\mathcal{Q}\mathcal{G}_0(0)\mathcal{Q}\mathcal{V}\mathcal{P}\rho
\end{split}
\label{eq:kernel}
\end{equation}
where the effective Liouvillian is \cite{Finkelstein2016-1}:
\begin{equation}
\begin{split}
L_{\text{eff}}&\equiv \mathcal{P}L_0\mathcal{P}+\mathcal{P}\mathcal{V}\mathcal{Q}\mathcal{G}_0(0)\mathcal{Q}\mathcal{V}\mathcal{P}\equiv \mathcal{P}L_0\mathcal{P}+\mathcal{W}
\end{split}
\label{eq:Partition-Liouvillian}
\end{equation}
where $L_0=\mathcal{P}L\mathcal{P}+\mathcal{Q}L\mathcal{Q}$ is the block-diagonal Liouvillian of the whole system, $\mathcal{G}_0(z)=(z-L_0)^{-1}$ is its resolvent and $\mathcal{V} = L-L_0$. The operator $\mathcal{W}=\mathcal{P}\mathcal{V}\mathcal{Q}\mathcal{G}_0\mathcal{Q}\mathcal{V}\mathcal{P}$ captures the effect of the continuum on the two-level system. The projection $\mathcal{P}\rho$ of the exact steady-state density matrix in the 4 dimensional subspace spanned by the discrete sates and their associated coherences, is obtained through the kernel of the effective Liouvillian which is a $4\times 4$ matrix. This is a considerable simplification as it can be easily done numerically or symbolically by the appropriate software.
In a second step, the population in the continuum set of states - the observable of interest - is obtained  from
Eq.~\eqref{eq:kernel} as \cite{Finkelstein2016-1}:
\begin{equation}
\begin{split}
\int dk \rho_{kk}&=tr(\mathcal{Q} \rho)=\sum_{\{ij\}} C_{\{ij\}} (\mathcal{P} \rho_0)_{\{ij\}},
\end{split}
\label{eq:Ci}
\end{equation}
where $C_{\{ij\}}=\int dk \bbra{kk} \mathcal{Q}\mathcal{G}\mathcal{Q}\mathcal{Q}L\mathcal{P}\kket{ij}$ with $\kket{ij}=\kket{gg},\kket{eg},\kket{ge},\kket{ee}$ and  $\mathcal{Q}\mathcal{G}\mathcal{Q}=(\mathcal{Q}(\underline{\Omega}_L-L)\mathcal{Q})^{-1}$. In this way, we have expressed the total population of the continuum as a linear combination of the elements of the density matrix in the subspace spanned by the discrete states only, with coefficients $C_{ij}$. Equations \eqref{eq:kernel} and \eqref{eq:Ci} give a method for calculating  the population in the continuum set of states. In general this is an involved calculation that does not always result in a closed form solution. We will show the conditions necessary for equation \eqref{eq:Ci} to conclude in a Fano profile, and will illustrate a simple method of evaluation of the effective operators in \eqref{eq:kernel} that highlights the connection between the scattering problem and the dissipative one.  \newline


\subsection{Conditions for a Beutler-Fano profile}
The Beutler-Fano lineshape characterizes the dependence of the continuum states population as a function of irradiation frequency, which appears in the model as the dimensionless parameter $\epsilon=(\omega_L-E_e)/\hbar\gamma$. To understand the origin of the Fano form we must understand the $\epsilon$-dependence of the population in the continuum. We note that the Beutler-Fano profile needs not to be exactly expressed as a function of $\epsilon$ but can be given as a function of an effective $\epsilon_{\text{eff}}$ that has been shifted and rescaled with respect to $\epsilon$, $\epsilon_{\text{eff}}= \frac{\epsilon+\Delta}{\sigma}$ (see appendix~\ref{app:FanoPolynome}). This will give the same functional lineshape.
It can be shown that a ratio of polynomials of order 2 in $\epsilon$ is equivalent to writing it as a Beutler-Fano lineshape plus a Lorentzian function, or alternatively as a Fano profile with a complex asymmetry parameter $\underbar{q} = q +iq_\text{i}$ (see Appendix~\ref{app:FanoPolynome}):
\begin{align}
f(\epsilon,q)&=\frac{\abs{\epsilon_{\text{eff}}+ \underbar{q}}^2}{\epsilon_{\text{eff}}^{2}+1}
=\frac{(q+\epsilon_{\text{eff}})^2}{\epsilon_{\text{eff}}^{2}+1} +  \frac{q_{\text{i}}^2}{\epsilon_{\text{eff}}^{2}+1} \nonumber \\
&=\frac{a_0 + a_1\epsilon + a_2\epsilon^2}{b_0 + b_1\epsilon + b_2\epsilon^2},
\end{align}
Therefore, to prove that the profile is of  Beutler-Fano type  is equivalent to prove that the observable is the ratio of two polynomials of order 2 in $\epsilon$. \newline

The effective Liouvillian is calculated by means of exact resummation of perturbative expansions as was carried out in Refs. \cite{Finkelstein2015} and \cite{Finkelstein2016-1}.
The result for the model that considers dissipation from the continuum states $\ket{k}$ and from the  discrete excited state $\ket{e}$ to the ground state $\ket{g}$ with dissipation rates $\Gamma_c$ and $\Gamma_e$ is:
\begin{equation}
L_{\text{eff}}=\left[\begin{matrix}
0 & K\Omega & K^*\Omega & 2\Gamma_e+2 \\
-K^*\Omega & A & 0 & -K\Omega \\
-K\Omega & 0 & A^* & -K^*\Omega \\
0 & -K\Omega & -K^*\Omega & -2
\end{matrix}\right]
\end{equation}
in units of $n\pi V^2$, where $K=1+iq$, $A=-\Gamma_e -\Omega^{2} - i \epsilon - \gamma_{eg} - 1$ and $\Omega=F\mu_c/2V$, $\gamma_{eg}$ is the dephasing rate of the two levels system (TLS) and all other parameters have been previously defined. We note that  the system is impervious to the pure dephasing processes between the continuum and the TLS in the wideband approximation, as . In preparation to tackling more general cases, we will  do two key steps:
\begin{enumerate}
\item \textit{Separate the effective Liouvillian in two parts: a scattering contribution and a generalized quantum jump operator} \\
\item \textit{Reduce the solution of the kernel of the $4\times 4$ $L_{\text{eff}}$ matrix to a linear equation of $3\times 3$ matrices using the conditions for relaxing maps (i.e. having a unique steady-state) \cite{Davies1970,Spohn1977} }
\end{enumerate}

For the first step we write:
\begin{equation}
L_{\text{eff}}=-i(1 \otimes H_{\text{eff}}-\bar{H}_{\text{eff}}\otimes 1)+\tilde{L}
\end{equation}
where we can recognize $H_{\text{eff}}$ as the effective Hamiltonian of the scattering problem Eq. \eqref{eq:Heff} and $\tilde{L}$ is a generalized quantum jump operator that ensures the conservation of the trace of the density matrix \cite{Plenio1998,Breuer2004,Gardiner1992,
Piilo2008}.
The quantum jump approach was originally introduced in the modelling of fluorescence decay. In our case $\tilde{L}$ not only restores population to the ground state from populations in the excited state, but also from ground-excited coherences as can be seen from the elements in the upper row of $\tilde{L}$. The total population in the continuum (Eq. \eqref{eq:Ci}) as a function of the populations of the disrete partition is fully specified by a column vector $\mathbf{C}$ containing the coefficients $C_i$ and an appropriate normalization $\int dk \rho_{kk}+\rho_{gg}+\rho_{ee}=1$. These three elements, $H_{\text{eff}}$, $\tilde{L}$, and $\mathbf{C}$ fully specify the solution:
\begin{equation}
\begin{split}
H_{\text{eff}}&=\begin{bmatrix}
-i\Omega^2 & (q-i)\Omega \\
(q-i)\Omega & -\epsilon-i \\
\end{bmatrix} \\
\tilde{L}&=\begin{bmatrix}
2\Omega^2 & 2\Omega & 2\Omega & 2 \\
0 & 0 & 0 & 0 \\
0 & 0 & 0 & 0 \\
0 & 0 & 0 & 0
\end{bmatrix}, \; \mathbf{C}=\begin{bmatrix}
2\Omega^2 \\ \Omega \\ \Omega \\ 1
\end{bmatrix}
\end{split}
\end{equation}

The second step involves calculating the kernel of $L_{\text{eff}}$ by transforming the problem to a linear equation $M v=b$. We can do this because the Lindblad form of a two-level system is written in the basis of Pauli matrices, so it fulfills the condition that the generators are self-adjoint and that only the identity commutes with them \cite{Spohn1977}. As a consequence there is a unique steady-state, one of the elements of the density matrix is determined by the normalization condition and the problem reduces to the linear equation of dimension $3 \times 3$. We use $\rho'$ to denote the unnormalized density matrix. Then:
\begin{equation}
\begin{split}
M&=\left[\begin{matrix}
0 & Q\Omega & Q^*\Omega \\
-Q^*\Omega & A & 0 \\
-Q\Omega & 0 & A^* \\
\end{matrix}\right];\;
v=\begin{bmatrix}
\rho_{gg}' \\
\rho_{ge}' \\
\rho_{eg}'
\end{bmatrix};\;
b= \begin{bmatrix}
-2\Gamma_e-2 \\
Q\Omega \\
Q^*\Omega
\end{bmatrix}
\end{split}
\label{eq:Cramer}
\end{equation}
Cramer's formula tells us that:
\begin{equation}
\begin{bmatrix}
\rho'_{gg} \\
\rho'_{eg} \\
\rho'_{ge} \\
\rho'_{ee}
\end{bmatrix}=\frac{1}{\text{det}(M)}\begin{bmatrix}
\text{det}(M_1) \\
\text{det}(M_2) \\
\text{det}(M_3) \\
\text{det}(M)
\end{bmatrix}
\end{equation}
where $\text{det}(M_i)$ is the determinant obtained from the matrix $M$ by replacing the $i$-th column by the  vector $b$. We can neglect the overall prefactor $\frac{1}{\text{det}(M)}$ since it will cancel during the normalization. We look at each element of the density matrix.
\begin{enumerate}
\item  The structure of the effective Liouvillian is such that the various elements $\text{det}(M_i)$ are polynomials in $\epsilon$ of order 0, 1 and 2 (see Equation \eqref{eq:Cramer}).
\item The population of the continuum is a linear combination of the density matrix in the $\mathcal{P}$ subspace with $\epsilon$-independent coefficients, so that along with the normalization condition $\rho_{gg}+\rho_{ee}+\int dk \rho_{kk}=1$ we arrive at the end result.
\end{enumerate}

The previous steps prove that
\begin{equation}
\int dk \rho_{kk}=\frac{\sum_{n=0}^2 a_n\epsilon^n}{\sum_{n=0}^2 b_n\epsilon^n}
\end{equation}
which by a judicious normalization of the variables can be brought back to a general Beutler-Fano profile (see Appendix~\ref{app:FanoPolynome}). The dependence on the square of $\epsilon$ arises from the determinant of $M_1$ that involves the product of the coherences. This is a direct consequence of the wideband approximation and can be traced back to the fact that $L_{\text{eff}}-L_0$ does not depend on $\epsilon$ (the same reasons lead to $\epsilon$-independent $C_i$ coefficients). After analyzing the possible generalizations of the dissipation channels we will revisit the structure of Eq. \eqref{eq:Cramer}. \newline

The condition that allows us to express the kernel problem as a linear equation is very general and stems from the assumption of a steady-state of dimension 1. This condition can be broken whenever the relaxation mechanism is towards a manifold of states that do not have dissipation within them and is not typical of most physical systems. In this case, we must solve for the kernel directly. 

It is also important to clarify the nature of the steady-state in particular in connection to photoelectron spectroscopy experiments in the presence of radiative damping \cite{Agarwal1984}. With pure Hamiltonian evolution, the asymptotic limit of the population will be entirely in the continuum. In the dissipative case where the generator of dynamics is a Lindblad operator there will always be a true steady-state which consists of a fraction of electrons in the ground state (which can be negligible if the dissipation rate is much smaller than the Rabi frequency). A more realistic description can be obtained by introducing a sink state which acts as a detector. Alternatively, if the atoms are continuously pumped into the system it can be seen as a transport experiment, a problem solved in the next section.

\subsection{Non-equilibrium stationary transport.}
\begin{figure}[h]
\centering
\begin{tikzpicture}[scale=0.6]
\draw[thick] (-4,0) -- (0cm,0) node[right]{$\ket{g}$};
\draw[thick] (0,3cm)--(4cm,3cm) node[right]{$\ket{e}$};
\draw[fill=gray]  (6cm,1cm) rectangle (7cm,5cm) node[right]{$\ket{k}$} ;
\draw[->,thick] (-2cm,0) --(2cm,2.9cm) node[midway, right] {$\mu_{e}$};
\draw [->,thick] (-1.5cm,0cm)--(6.25cm,3.25cm)
node[midway, above] {$\mu_{k}$};
\draw[<->,thick] (3cm,3.1cm) to[out=45,in=135] node [sloped, above] {$V(k)$} (6.5cm,3.1cm);
\draw [->,thick,decorate,decoration=snake] (1cm,2.9cm)--(-3cm,0) node[midway,left] {$\Gamma_e$};
\draw[->,line width=1.5] (7.1cm,2.5) -- (8cm,2.5);
\draw[line width=1.5] (8cm,2.5) -- (9cm,2.5);
\draw[line width=1.5] (9cm,2.5) -- (9cm,-1cm);
\draw[->,line width=1.5] (9cm,-1cm) -- (2cm,-1cm) node[pos=0.5,anchor=south]{$J =\int dk \Gamma_c \rho_{kk}=r\rho_{gg}$};
\draw[line width=1.5] (2cm,-1cm) -- (-5cm,-1cm);
\draw[line width=1.5] (-5cm,-1cm) -- (-5cm,0cm);
\draw[->,line width=1.5] (-5cm,0cm) -- (-4.1cm,0cm);
\end{tikzpicture}
\caption{\label{fig:transportFano} Energy levels and transitions of a Fano-type model with dissipation for the electronic transport set-up. Hamiltonian coupling are indicated by straight arrows, dissipative processes by curly arrows. Population relaxations  from the discrete excited state, at a rate $\Gamma_e$ to the ground is considered. The population relaxation from the continuum set of states is now replaced by as a stationary current $J$ flowing across the systems.}
\end{figure}
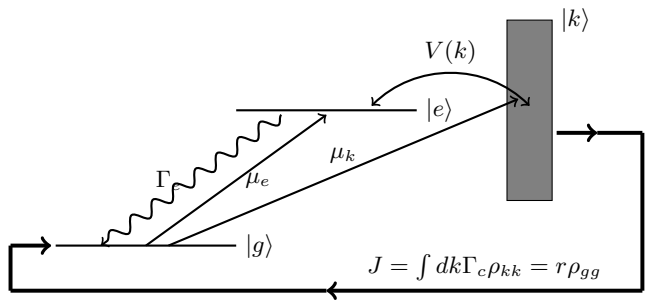

Beutler-Fano profiles has also been predicted and observed in electronic transport \cite{Huang2015,Baernthaler2010,Xiao2016} where a confined electronic quantum system (atom, molecule, quantum dot, circuit) is connected to electrodes.
As noted in Refs~\cite{Nitzan2002,Weiss2006} the knowledge of the equilibrium stationary state $\rho$, obtained in the previous section through Eq.~\eqref{eq:Partition-Liouvillian} and Eq.~\eqref{eq:Ci} can also be used to describe the following non-equilibrium stationary electronic transport situation.
Take the same system as above but where population from the continuum state to the ground state, at rate $\Gamma_c$,  is now replaced by a current $J$ flowing across the system,  with $J =\int dk \Gamma_c \rho_{kk}$ (see figure~\ref{fig:transportFano}).
The master equation describing this transport set-up can be written as:
\begin{equation}
\dot{\rho}=L_{\text{t}}\rho+J,
\label{eq:FanoTransport}
\end{equation}
where $L_{\text{t}}=L-\int dk \Gamma(k) A(k,g)\otimes A(k,g)$ is the same Liouvillian as the original $L$ (see Eq.~\eqref{eq:dissipation_k}), except that it does not include the  population relaxation from the continuum set of states back to the ground state.
It was shown that the stationary solution of Eq.~\eqref{eq:FanoTransport} is the same as the equilibrium stationary solution $L\rho=0$, if the current $J$ is exactly taken as $J =\int dk \Gamma(k) \rho_{kk}$ \cite{Nitzan2002,Weiss2006}. This was carried out in the context of electron transport through a molecular bridge.
Moreover,  because the current is stationary, it can also be written as $J = r\rho_{gg}$ where the transfer rate $r$ is from the ground state to any of the continuum states. This transfer rate can therefore be expressed as:
\begin{equation}
r = \frac{\int dk \Gamma(k) \rho_{kk}}{\rho_{gg}}.
\label{eq:Transport}
\end{equation}
The transfer rate $r$ is an intrinsic characteristic of the system coupled to the laser field.
It is independent of $\Gamma(k)$ which is a constant noted $\Gamma_c$ in the wide band approximation. The constant $\Gamma_c$ only sets the time scale to reach the stationary regime.
The transfer rate $r$ gives the probability per unit of time for an electron to  jump from the ground state to one of the continuum states.
We have shown in Ref.~\cite{Finkelstein2016-1} that $r$ as a function of $\epsilon$ can also be written as the sum of a Beutler-Fano profile and a Lorentzian function. Furthermore, in the limit of low field, with $\Omega\ll 1$ and setting $\Gamma_e = 0$ (that is ignoring any dissipative process) we exactly recover the scattering rate
given by Eq.~\eqref{eq:scattRate}~:
\[
r(\epsilon;\Gamma_e=0) = \frac{\ud P}{\ud t} + \mathcal{O}(\Omega^4)= 2\Omega^2\frac{(\epsilon +q)^2}{1+\epsilon^2} +
\mathcal{O}(\Omega^4).
\]
Consequently, we can affirm  that this transport set-up constitutes a well defined generalization of the original Fano scattering formalism, adapted to include dissipative processes and intense incident laser field.

We have explicitly proved that in a very general way a Beutler-Fano profile is observed even for intense incident fields and when dissipative processes are included. Doing so, we have also developed a method to obtain the stationary state for a dissipative system with continuum spectrum, which relies on obtaining an effective Liouville operator in the discrete states subspace, through partitioning and resummation of Dyson equations.
In the next section, we will use this method to tackle the more general problems of multiple discrete levels coupled to multiple continua with arbitrary Markovian dissipative channels. In theses cases, we expect to observe a departure from the strict Beutler-Fano profile.

The model presented in the last section is an example of a discrete-continuum Hamiltonian with Markovian dissipation channels, but both the Hamiltonian structure and Markovian channels can be generalized. We start by reviewing the possible additional dissipation channels and then move on to consider an arbitrary discrete-continuum Hamiltonian.

\subsection{Generalizing the dissipation: incoherent hopping and finite temperature effects}
  A first extension is to include  an incoherent decay at a rate $\Gamma_{ce}$ from the continuum states to the discrete excited state in addition to the decay to the ground state at a rate $\Gamma_{cg}$.
The effective Liouvillian can be obtained following the same technique as before.
For ease of readability, in this section and what follows the explicit form of the effective Liouvillians will be listed in the Appendix~\ref{app:FormOperators}. In this case $L_{\text{eff}}-L_0$ is also $\epsilon$-independent and the system relaxes to the ground state (due to $e-g$ population decay), so that we may say that the final profile will be in Fano form as well. We can follow a similar line of reasoning as for the preceding section with the use of Cramer's rule. Here the general $M$ matrix is:
\begin{equation}
\begin{split}
M&=\left[\begin{matrix}
K & C & C^* \\
B^* & A & 0 \\
B & 0 & A^* \\
\end{matrix}\right]
\end{split}
\end{equation}
where $K=2\Omega^2(\beta-1)$, with $\beta=\frac{\Gamma_{cg}}{\Gamma_{cg}+\Gamma_{ec}}$; and  $B=-\Omega(1+iq)$,  $C=\Omega(\beta-1+iq)$. The same order of $\epsilon$ dependences indeed lead to a Fano profile plus a Lorentzian. \newline


Finite temperature effects, or incoherent terms that will take population from the discrete states and transfer it to the continuum partition are also physically important. Let's consider the rate from ground state to the continuum $\Gamma_{g\to k}$. The total injection into the continuum $\int dk \Gamma_{g\to k}$ diverges in the wideband approximation. It is a problem that only exists when the accepting states are not finite. This is resolved because physically the continuum does not extend to infinity. This means that incoherent pathways into the continuum cannot be included within the wideband approximation and go beyond the ambitions of this work. Extending beyond the wideband approximation in Liouville space is considerably more complicated than in the Hilbert space and will be presented in the future. \newline

Once having established the possible most general relaxation channels, we move on to generalize the solution of Fano interference in Hamiltonians of arbitrary complexity, meaning multiple discrete levels and multiple continua where the wideband approximation is still valid.

\subsection{Multiple discrete-continuum Hamiltonian with Markovian dissipation channels}

In his seminal 1961 article \cite{Fano1961}, these multi-levels multi-continua structures were addressed but in  the weak-field Hamiltonian scattering approach of Section I., but to our knowledge, no general solution have been published to date in Liouville space.
As we will see, while their solution does not conform to the Beutler-Fano profile they can be worked out by the effective Liouvillian approach explained in this article.
The method of solution is similar as to what has been presented before and we provide a recipe to calculate the continuum population in the most general case.

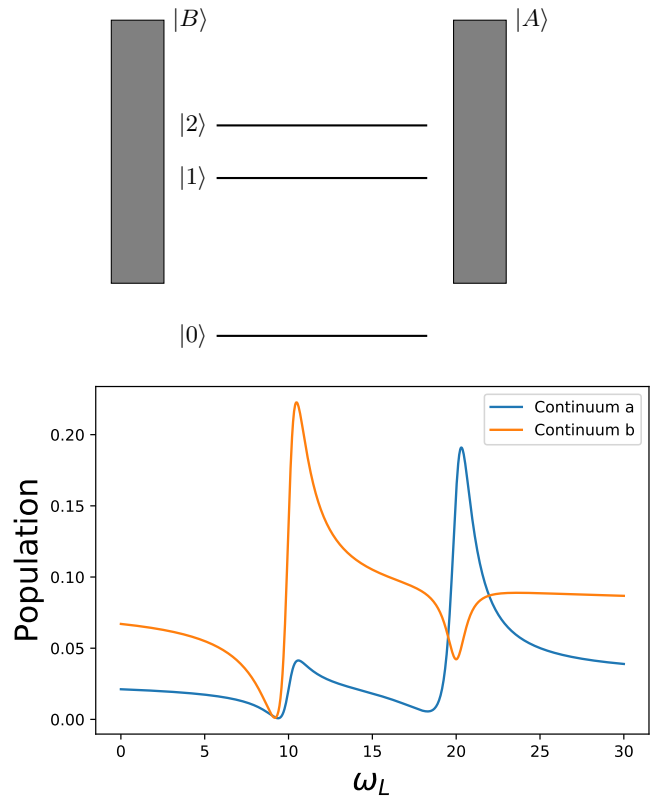
\begin{figure}[ht]
\centering$
\begin{array}{cc}
\begin{tikzpicture}[scale=0.7]
\draw[thick] (0,0) node[left]{$\ket{0}$} -- (4cm,0);
\draw[thick] (0,3cm) node[left]{$\ket{1}$} --(4cm,3cm);
\draw[thick] (0,4cm) node[left]{$\ket{2}$} --(4cm,4cm);
\draw[fill=gray]  (4.5cm,1cm) rectangle (5.5cm,6cm) node[right]{$\ket{A}$} ;
\draw[fill=gray]  (-2cm,1cm) rectangle (-1cm,6cm) node[right]{$\ket{B}$} ;
\end{tikzpicture} \\ \includegraphics[width=0.5\textwidth]{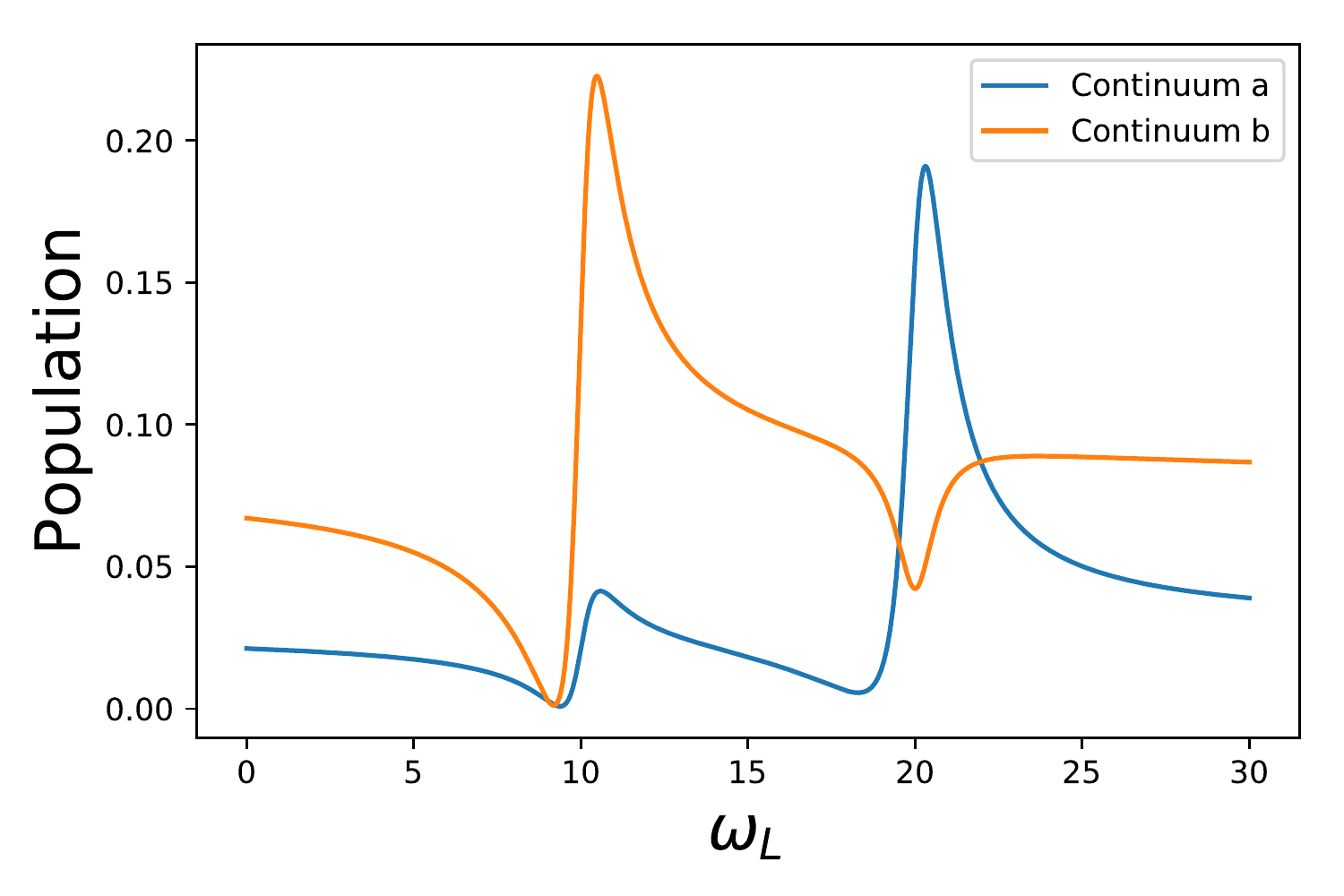}
\end{array}$
\caption{\label{fig:multiple_continua} Populations in a general Fano system consisting of three discrete levels ($0$, $1$, and $2$) and two continua $A$ and $B$. The parameters and couplings are $E_1=10$, $E_2=20$, $\mu_{01}=0.3$, $\mu_{02}=0.4$, $V_{12}=0$, $\Gamma_{31}=0.05$, $\Gamma_{21}=0.04$, $V_{1}^{(A)}=0.05$, $V_{2}^{(A)}=0.1$, $V_{3}^{(A)}=0.2$, $V_{1}^{(B)}=0.1$, $V_{2}^{(B)}=0.3$, $V_{3}^{(B)}=0.02$, $\Gamma^{(A)} =0.5$, $\Gamma^{(B)}=0.7$. $\mu_{ij}$ is the transition dipole moment between states $i$ and $j$, $V_{ij}$ the electronic coupling and $\Gamma$ the relaxation rates.}
\end{figure}

We obtain the solution by first calculating the effective Liouvillian in the discrete state subspace. We consider an $N$-levels system coupled to $M$ separate continua.  The continuum $a$ will relax to the discrete level $b$ with a rate $\Gamma^{(a)}_{b}$. We have:
\begin{equation}
\begin{split}
H_0&=\sum_{i=1}^N E_i \ket{i}\bra{i} + \sum_{\substack{i,j=1 \\ i\neq j}}^N \mu_{ij} \ket{i}\bra{j} \\
H_{\text{eff}}-H_0&=-i\sum_{a=1}^M\sum_{i,j=1}^{N}n^{(a)}\pi V_{i}^{(a)}V_{j}^{(a)}\ket{i}\bra{j} \\
L_D&=\sum_k \Gamma_k\big(  D_k\rho D_k^{\dagger}-\frac{1}{2}\{D_k^{\dagger}D_k,\rho \}\big)\\
\tilde{L}&=\sum_{a=1}^{M}\sum_{i,j=1}^{N}\sum_{b=1}^{N}\frac{2\Gamma_{b}^{(a)}}{\sum_{l=1}^{N}\Gamma_{l}^{(a)}}
n^{(a)} \pi V_{i}^{(a)}V_{j}^{(a)}\kket{bb}\bbra{ij} \\
C^{(a)}_{\{ij\}}&= \sum_{i,j=1}^{N} \frac{n^{(a)}2\pi V_{i}^{(a)}V_{j}^{(a)}}{\sum_{l=1}^{N}\Gamma_{l}^{(a)}}  \\
& \sum_{a=1}^{M} \int dk_a \rho_{k_ak_a} + \sum_{b=1}^{N} \rho_{bb} = 1
\end{split}
\label{eq:recipe}
\end{equation}
where $V_i^{(a)}$ is the coupling (including radiative coupling) between level $i$ and the continuum $(a)$ and
$\Gamma_b^{(a)}$ is the relaxation rate from the continuum $a$ to the discrete state $b$. $L_D$ is the dissipation within the discrete manifold and here $D_i$ are the Krauss operators corresponding to the discrete manifold only. The transformation into the dimensionless constants appearing throughout this article is straightforwardly obtained by normalizing by the effective width of choice $\gamma=n\pi V^2$ (where the chosen coupling $V$ varies depending on the Hamiltonian structure).
This is one of the main results of this article and a landmark result in the Fano literature. It is the generalization of the Fano problem to any discrete-continuum Hamitlonian in a Markovian environment at zero temperature within the wideband approximation. Figure \ref{fig:multiple_continua} shows the population in the continuum of a general Fano system consisting of three discrete levels ($0$, $1$, and $2$) and two continua $A$ and $B$ along with the characteristic asymmetric lineshapes.


We summarize the difference between the approach followed in this article and in particular the one followed in the papers by Agarwal and co-workers \cite{Agarwal1984}. A fundamental difference stems from the construction of the dissipation superoperator which is tied to the radiation field in the case of Agarwal et al. and is purely phenomenological in the case presented. Both suffer from the local approximation which requires a reinterpretation of the value of the dissipation rates as the values of the Hamiltonian couplings are varied. The apporaches to a solution also differ. In \cite{Agarwal1984}, the solution is obtained by an elegant density factorization matrix that gives the time-evolution of the populations. In this case (and our other works \cite{Finkelstein2016-1}), we use Feshbach projection methods that solve directly for the steady-state by working in the superoperator space. The advantage is that the extension to any number of arbitrary continua and discrete states is straightforward, both populations and coherences are simultaneously solved. The dynamics is also obtaineable from this approach and will be addressed in subsequent work. A disadvantage worth noting is that deviations of the wideband approximation (for example an energy dependent dissipation rate) can only be solved perturbatively while in \cite{Agarwal1984} some energy dependences can be solved exactly.

\section{Conclusion}

The celebrated Fano profile describes a phenomenological dependence on the irradiation wavelength that is common to many theories. Often times the apparent simplicity of this form makes it easy to forget the host of phenomena that unfolds as the field becomes more intense. This behavior is strongly dependent on the experimental system and configuration of the experiment, and dictates the fundamentally important decision to use Hilbert space or Liouville space descriptions. The results are quite different. The Liouville space solution is solved via an effective superoperator method that reduces the problem of taking the kernel of an infinite matrix to taking the kernel of a $4 \times 4$ matrix. We have analyzed this structure in detail and shown the mathematical arguments to obtain a Fano profile. We have also generalized the approach to any discrete-continuum Hamiltonian coupled to a Markovian bath under the wideband approximation. Population in the continua of these systems can be now straightforwardly obtained (see Figure \ref{fig:multiple_continua}).

\section{Acknowledgements}

D.F.S. thanks a Marie-Sklodowska-Curie Individual Fellowship. We thank O. Atabek and V. Mujica for fruitful discussion.

\clearpage

\appendix
\section{Projections}
\label{app:Proj}
\textbf{Hilbert space}. We can project Dyson equation onto the discrete and continuous states using $P$ and $Q$  operators so that $G=G_0+G_0VG=G_0+GVG_0$ becomes:

\begin{equation}
\begin{split}
&{P}{G}{P}={P}{G}_0{P}+{P}{G}_0{P}({P}{V}{Q}{G}_0{Q}{V}{P}){P}{G}{P} \\
&{Q}{G}{P}={Q}{G}_0{Q}{V}{P}{G}{P} \\
&{P}{G}{Q}={P}{G}{P}{P}{V}{Q}{Q}{G}_0{Q} \\
&{Q}{G}{Q}={Q}{G}_0{Q}+{Q}{G}_0{Q}({Q}{V}{P}{G}{P}{V}{Q}){Q}{G}_0{Q}
\end{split}
\end{equation}
Multiplying the first equation by $z-H_0$, we obtain~
$\left(z-H_{\text{eff}}\right){P}{G}{P}=\un$ where $H_{\text{eff}}=PH_0P+PVQG_0QVP$.
In this way, we can compute the projection of the exact resolvent in the $P$ subspace as the resolvent of $H_{\text{eff}}$,  an operator that acts only in this subspace.
\newline

\textbf{Liouville space.} Here the projection superoperators are defined as:
\begin{equation}
\mathcal{P}=P\otimes P;\quad%
\mathcal{Q}=P\otimes Q + Q\otimes P + Q\otimes Q.
\label{eq:partitionsL}
\end{equation}
and the same idea can be applied. Interested in the steady-state we concentrate on the kernel of the Liouvillian $L\rho=0$. Inserting the identity $\mathcal{P} + \mathcal{Q}$, and projecting in each subspace, yields:
\begin{align}
\mathcal{P}L\mathcal{P}\rho+\mathcal{P}L\mathcal{Q}\rho&=0 \\
\mathcal{Q}L\mathcal{P}\rho+\mathcal{Q}L\mathcal{Q}\rho&=0
\label{eq:PQL}
\end{align}
We define $L_0 = \mathcal{P}L\mathcal{P}$ + $\mathcal{Q}L\mathcal{Q}$ and
$\mathcal{V} = L-L_0$.
Multiplying the second line of Eq.~\eqref{eq:PQL} by $\mathcal{Q}\mathcal{G}_0(z)\mathcal{Q}$ where:
\[
\mathcal{Q}\mathcal{G}_0(z)\mathcal{Q} = \mathcal{Q}(z-L_0)^{-1}\mathcal{Q} =
(z-\mathcal{Q}L_0\mathcal{Q})^{-1}
\]
and taking the limit $z=0$, yields
\[
\mathcal{Q}\rho = \mathcal{Q}\mathcal{G}_0(0)\mathcal{Q}L\mathcal{P} \rho = \mathcal{Q}\mathcal{G}_0(0)\mathcal{Q}\mathcal{V}\mathcal{P} \rho
\]
Inserting this last equation in the second term of the first line of Eq.~\eqref{eq:PQL}
give the expressions in the main text.
\section{Lineshape as a quotient of polynomials}
\label{app:FanoPolynome}
We show that there is a one-to-one correspondence between a generalized Fano profile and a quotient of polynomials of order 2 in the laser wavelength, or the parameter $\epsilon=(\omega_L-E_e)/n\pi V^2$. For now, we assume that the population in the continuum $n_c$ can be written as:
\begin{equation}
\int dk \rho_{kk}=\frac{\sum_{n=0}^2 a_n\epsilon^n}{\sum_{n=0}^2 b_n\epsilon^n}
\end{equation}
We rework these expressions to show that the above expression is equivalent to a Fano plus a Lorentzian. We begin by rescaling the parameter $\epsilon$ so that $\epsilon'=(\epsilon+\Delta)/\sigma$ with $\Delta=b_1/2b_2$ and $\sigma=\sqrt{b_0/b_2-b_1^2/4b_2^2}$. Then,
\begin{equation}
\textbf{denominator}=K(\epsilon'^2+1)
\end{equation}
where $K=b_0-b_1^2/4b_2$. The denominator by itself describes  a Lorentzian shifted in resonance from $\omega_L$ by $\Delta$ and further broadened by a factor $\sigma$.

The population of the continuum now reads:
\begin{equation}
\int \rho_{kk} dk=\frac{\sum_{n=0}^2 a_n\epsilon^n}{K(\epsilon'^2+1)}
\end{equation}
Given that we have rescaled the denominator, we now work on the numerator which we write as
\begin{equation}
\begin{split}
\textbf{numerator}&=\sum_{n=0}^2a_n\epsilon^n \\
&=\sum_{n=0}^2a_n(\sigma\epsilon'-\Delta)^n \\
&\equiv \sum_{n=0}^2c_n(\epsilon')^n \\
&=c_2( (\epsilon'+q)^2+D )
\end{split}
\end{equation}
where we have defined $c_2=a_2\sigma^2$, $c_1=a_1\sigma -2a_2\sigma\Delta$, $c_0=a_0-a_1\Delta$, with $q=c_1/2c_2$ and $D=c_0/c_2-c_1^2/4c_2^2$. This last form corresponds, along with a denominator, to a Fano profile plus a Lorentzian. Thus we see that there is a one-to-one correspondence between a Fano plus Lorentzian term and the quotient of two polynomials of order 2. We now have to show that the population of the continuum is a ratio of two polynomials of order 2.

\section{The wideband approximation in Liouville space}
The wideband approximation takes on different forms in Hilbert and Liouville space. In Hilbert space projecting out the continuum involves one integral whose principal part vanishes in the wideband approximation. In Liouville space projecting out the continuum involves an infinity of integrals that are products of simple poles. These poles always lie on one side of the real axis, so that the wideband approximation allows to draw a contour in a semi-infinite plane resulting in all of the integrals containing more than one pole vanishing, and all of the rest with one pole evaluating to the energy independent value of $-in\pi$.

\section{Explicit form of the operators}
\label{app:FormOperators}
\textbf{General dissipation.} The general form of the dissipation is:
\begin{equation}
H_{\text{eff}}=\begin{bmatrix}
-i\Omega^2 & (q-i)\Omega \\
(q-i)\Omega & -\epsilon-i \\
\end{bmatrix}
\end{equation}
\begin{equation}
\begin{split}
L_{QJ}&=\beta \begin{bmatrix}
2\Omega^2 & 2\Omega & 2\Omega & 2 \\
0 & 0 & 0 & 0 \\
0 & 0 & 0 & 0 \\
0 & 0 & 0 & 0 \\
\end{bmatrix} + (1-\beta) \begin{bmatrix}
0 & 0 & 0 & 0 \\
0 & 0 & 0 & 0 \\
0 & 0 & 0 & 0 \\
2\Omega^2 & 2\Omega & 2\Omega & 2 \\
\end{bmatrix}
\end{split}
\end{equation}
where $\beta=\frac{\Gamma_{cg}}{\Gamma_{cg}+\Gamma_{ce}}$.

We can follow a similar line of reasoning as for the preceding section with the use of Cramer's rule. Here the general $M$ matrix is:
\begin{equation}
\begin{split}
M&=\left[\begin{matrix}
K & C & C^* \\
B^* & A & 0 \\
B & 0 & A^* \\
\end{matrix}\right]
\end{split}
\end{equation}
which gives a determinant $det(M)$ which does depend on $\epsilon$ so that the final expression now includes higher order terms and can in principle not be expressed as a Fano lineshape any longer. \newline

\textbf{Multiple discrete levels coupled to a continuum.} We fully specify the solution for the case of 3 discrete states coupled to one continuum:
\begin{equation}
H_{\text{eff}}=\begin{bmatrix}
-i\Omega^2 & (q_1-i)\Omega & \frac{(q_2-i)\Omega}{\beta} \\
(q_1-i)\Omega & -\epsilon-i & -\frac{i}{\beta} \\
 \frac{(q_2-i)\Omega}{\beta} & -\frac{i}{\beta} & -\epsilon+\delta-\frac{i}{\beta^2}
\end{bmatrix}
\end{equation}
\begin{equation}
\begin{split}
\tilde{L}&=2\begin{bmatrix}
\Omega^2 & \Omega & \Omega/\beta & \Omega & 1 & 1/\beta & \Omega/\beta & 1/\beta & 1/\beta^2  \\
0 & 0 & 0 & 0 & 0 & 0 & 0 & 0 & 0 \\
0 & 0 & 0 & 0 & 0 & 0 & 0 & 0 & 0 \\
0 & 0 & 0 & 0 & 0 & 0 & 0 & 0 & 0 \\
0 & 0 & 0 & 0 & 0 & 0 & 0 & 0 & 0 \\
0 & 0 & 0 & 0 & 0 & 0 & 0 & 0 & 0 \\
0 & 0 & 0 & 0 & 0 & 0 & 0 & 0 & 0 \\
0 & 0 & 0 & 0 & 0 & 0 & 0 & 0 & 0 \\
0 & 0 & 0 & 0 & 0 & 0 & 0 & 0 & 0 \\
\end{bmatrix} \\
\int dk \rho_{kk}&=\frac{2}{\Gamma_c}\bigg[\Omega^2 \rho_{gg}+ 2 \Re \left( \Omega\rho_{ge_1})+ \frac{\Omega}{\beta} \rho_{ge_2}+\frac{1}{\beta}\rho{e_1e_2}\right) \\
&+ \rho_{e_1e_1}+\frac{1}{\beta^2}\rho_{e_2e_2} \bigg]
\end{split}
\end{equation}
where $\beta = \frac{V_1}{V_2}$, $\delta=(E_2-E_1)/n\pi V_1^2$ and the density matrix is subject to the appropriate normalization condition. \newline

\textbf{Multiple continua for one level.} In the case where the discrete excited state is coupled to more than one continuum, the effective Liouvillian is written as:

\begin{equation}
H_{\text{eff}}=\begin{bmatrix}
-i\sum_n \gamma_n^2\Omega_n^2 & q-i\sum_n\gamma_n^2\Omega_n \\
q-i\sum_n\gamma_n^2\Omega_n & -\epsilon-i \\
\end{bmatrix}
\end{equation}

\begin{equation}
\begin{split}
\tilde{L}&=\sum_{n=1}^2 \gamma_n^2 \begin{bmatrix}
2\Omega_n^2 & 2\Omega_n & 2\Omega_n & 2 \\
0 & 0 & 0 & 0 \\
0 & 0 & 0 & 0 \\
0 & 0 & 0 & 0 \\
\end{bmatrix},\;\; C_n=\frac{2\gamma_n^2}{\Gamma_{cn}}\begin{bmatrix}
\Omega_n^2 \\ \Omega_n \\ \Omega_n \\ 1
\end{bmatrix}
\end{split}
\end{equation}
where $\gamma_i^2=\frac{V_i^2}{V_1^2+V_2^2}$, with $\sum_n \gamma_n^2=1$.
The effective Liouvillian can almost be written as the one for a single continua except that the decay term $\sum_n \gamma_n^2\Omega_n^2$ is not the square of the off-diagonal elements $\sum_n \gamma_n^2\Omega_n$. The structure of the Liouvillian is such that the determinant of the sub-matrix (see preceding section) does not depend on $\epsilon$ so that the functional form is still in Fano form.
\clearpage

\bibliography{./Fano}

\end{document}